\documentclass{elsart}   

\usepackage{graphicx}
\usepackage{epsfig}

\def\alpha{\Greekmath 010B }%
\def\beta{\Greekmath 010C }%
\def\gamma{\Greekmath 010D }%
\def\delta{\Greekmath 010E }%
\def\epsilon{\Greekmath 010F }%
\def\zeta{\Greekmath 0110 }%
\def\eta{\Greekmath 0111 }%
\def\theta{\Greekmath 0112 }%
\def\iota{\Greekmath 0113 }%
\def\kappa{\Greekmath 0114 }%
\def\lambda{\Greekmath 0115 }%
\def\mu{\Greekmath 0116 }%
\def\nu{\Greekmath 0117 }%
\def\xi{\Greekmath 0118 }%
\def\pi{\Greekmath 0119 }%
\def\rho{\Greekmath 011A }%
\def\sigma{\Greekmath 011B }%
\def\tau{\Greekmath 011C }%
\def\upsilon{\Greekmath 011D }%
\def\phi{\Greekmath 011E }%
\def\chi{\Greekmath 011F }%
\def\psi{\Greekmath 0120 }%
\def\omega{\Greekmath 0121 }%
\def\varepsilon{\Greekmath 0122 }%
\def\vartheta{\Greekmath 0123 }%
\def\varpi{\Greekmath 0124 }%
\def\varrho{\Greekmath 0125 }%
\def\varsigma{\Greekmath 0126 }%
\def\varphi{\Greekmath 0127 }%

\def\nabla{\Greekmath 0272 }
\def\FindBoldGroup{%
   {\setbox0=\hbox{$\mathbf{x\global\edef\theboldgroup{\the\mathgroup}}$}}%
}

\def\Greekmath#1#2#3#4{%
    \if@compatibility
        \ifnum\mathgroup=\symbold
           \mathchoice{\mbox{\boldmath$\displaystyle\mathchar"#1#2#3#4$}}%
                      {\mbox{\boldmath$\textstyle\mathchar"#1#2#3#4$}}%
                      {\mbox{\boldmath$\scriptstyle\mathchar"#1#2#3#4$}}%
                      {\mbox{\boldmath$\scriptscriptstyle\mathchar"#1#2#3#4$}}%
        \else
           \mathchar"#1#2#3#4%
        \fi 
    \else 
        \FindBoldGroup
        \ifnum\mathgroup=\theboldgroup 
           \mathchoice{\mbox{\boldmath$\displaystyle\mathchar"#1#2#3#4$}}%
                      {\mbox{\boldmath$\textstyle\mathchar"#1#2#3#4$}}%
                      {\mbox{\boldmath$\scriptstyle\mathchar"#1#2#3#4$}}%
                      {\mbox{\boldmath$\scriptscriptstyle\mathchar"#1#2#3#4$}}%
        \else
           \mathchar"#1#2#3#4%
        \fi                 
          \fi}

\newif\ifGreekBold  \GreekBoldfalse
\let\SAVEPBF=\pbf
\def\pbf{\GreekBoldtrue\SAVEPBF}%
\begin{document}

\begin{frontmatter}

\title{Diffusion on Complex Networks : 
    A way to probe their large scale topological structures} 

\author[NTNU,NORDITA]{Ingve Simonsen}
\author[NORDITA,Lund]{Kasper Astrup Eriksen} 
\author[BNL]{Sergei Maslov} 
\author[NORDITA,NBI]{Kim Sneppen} 

\address[NTNU]{Department of Physics, NTNU, NO-7491, Trondheim, Norway}
\address[NORDITA]{NORDITA, Blegdamsvej 17, DK-2100~Copenhagen {\O}, Denmark}
\address[Lund]{Department of Theoretical Physics, Lund University,
S{\"o}lvegatan 14A, SE-223~62~Lund, Sweden}
\address[BNL]{Department of Physics, Brookhaven National Laboratory,
         Upton,~New~York~11973,  USA}
\address[NBI]{The Niels Bohr Institute,
Blegdamsvej 17, DK-2100 Copenhagen {\O}, Denmark}

\date{\today}


\begin{abstract}
  A diffusion process on complex networks is introduced in order to
  uncover their large scale topological structures.  This is achieved
  by focusing on the slowest decaying diffusive modes of the network.
  The proposed procedure is applied to real-world networks like a
  friendship network of known modular structure, and an Internet
  routing network. For the friendship network, its known structure is
  well reproduced. In case of the Internet, where the structure is far
  less well-known, one indeed finds a modular structure, and  modules
  can roughly be associated with individual countries. Quantitatively
  the modular structure of the Internet manifests itself in an
  approximately $10$ times larger participation ratio of its slowest
  decaying modes as compared to the null model -- a random scale-free
  network. The extreme edges of the Internet are found to correspond
  to Russian and US military sites.
\end{abstract}


\end{frontmatter}


\section{Introduction}

Complex networks are natural structures for representing many
real-world systems. Their flexible and adaptive features make them
ideal for interrelating various types of information and to update them
as time goes by.  Recently, there has been tremendous interest, from
various fields of science, in the study of the statistical properties
of such networks~\cite{Albert2002,Newman2003}.  By now, many
interesting features of complex networks have been established.
So far, the main attention in the study of complex networks has been
on the properties of individual nodes and how they connect (link) to
their nearest neighbors. However, exploring the {\em local} properties
of the network outside the nearest neighbor level, has not been well
studied. A noteworthy exception is the recent study by Girvan and
Newman~\cite{Girvan2002}.  Such studies will naturally have to address
the cluster (modular) structure of the network. To know if a network
is modular or not, is important if one tries to assess its robustness
and stability, since only a few critical links are responsible for the
inter-modular communication in clustered networks.  Hence to make the
network more robust, such critical links have to be identified and
strengthen.

In this work we will address the large scale topological structures of
networks. This problem will be approached by considering an auxiliary
diffusion process on the underlying complex network. As it will turn
out, it is the slowest decaying diffusive eigenmodes that will be of
most interest to us, since such modes will contain, as we will
see, information about the weakly interacting modules of the network.
Hence, by studying how an ensemble of random walkers slowly reaches a
state of equilibrium, one can learn something about the topology of the
network. 

To study spectral properties of networks is not new; Its variants has
previously been applied to random graphs~\cite{RandomGraphs}, to
social networks (the correspondence analysis)~\cite{wasserman_book},
random and small-world networks (the Laplace equation
analysis)~\cite{monasson}, artificial scale-free
networks~\cite{FarkasBarabasiSpectra,GohKahngKimSpectra}, and
community structures~\cite{Girvan2002,Kleinberg}. A diffusion approach
has also made it to practical applications.  For instance, the
analysis of a diffusion process lies at the heart of the
popular search engine Google \cite{google}. 


\section{A Markovian diffusion process on complex networks}
\label{Sec:Method}

The original physical motivation behind the method to be outlined
below, was that the relaxation of some arbitrary initial state (of
walkers on the network) toward the steady state distribution, was
expected to be {\em fast} in regions that are highly connected, while
{\em slow} in regions that had low connectivity.  By definition, a
module is highly connected internally, but has only a small number of
links to nodes outside the module. Hence, it was reasoned, that by
identifying the slowly decaying eigenmodes of the diffusive network
process, one should be able to obtain information about the large scale
topological structures, like modules, of the underlying complex
network. We will now formalize this idea and outline the method in
some detail. 

Let us start by assuming that we are dealing with a (fully connected)
complex network consisting of $N$ nodes and characterized by some
degree distribution $p(k)$ where $k_i$ denotes the connectivity of
node $i$.  Imagine now placing a large number ($\gg N$) of random
walkers onto this network.  The fraction of walkers on node $i$
(relative the total number) at time $t$ we will denote by $\rho_i(t)$.
In each time step, the walkers are allowed to move randomly between
nodes that are directly linked to each other. Since there is no way
that a walker can vanish from the network, the total number of walkers
must be conserved globally, {\it i.e.} one must have $\sum_i
\rho_i(t)=1$ at all times.  Furthermore, locally a continuity equation
must be satisfied; If we consider, say node $i$, that directly links
to other nodes $j$, one must require that (the balance equation):
\begin{eqnarray}
  \label{Eq:Cont}
  \rho_i(t+1)-\rho_i(t) &=& \sum_j A_{ij} \frac{\rho_j(t)}{k_j}
         - \sum_j A_{ji} \frac{\rho_i(t)}{k_i} .
\end{eqnarray}
In writing this equation, we have introduced the so-called adjacency
matrix $A_{ij}=A_{ji}$ defined to be $1$ if node $i$ and $j$ are
directly linked to each other, and $0$
otherwise~\cite{Albert2002,Newman2003}, and $k_{i}$, we recall, is the
degree of node $i$, {\it i.e.} its number of nearest neighbors $k_{i}
= \sum_{j} A_{ij}$.  The first term on the right hand side of
Eq.~(\ref{Eq:Cont}) describes the flow of walkers into node $i$, while
the last term is associated with the out-flow of walkers from the same
node.  As will be useful later, Eq.~(\ref{Eq:Cont}) can be casted into
the following equivalent matrix form:
\begin{eqnarray}
  \label{Eq:diffusion}
  \partial_t {\mathbf \rho}(t) &=& {\mathbf D} {\mathbf\rho}(t),
\end{eqnarray}
where $\partial_t {\mathbf \rho}(t) = {\mathbf \rho}(t+1) -{\mathbf
  \rho}(t)$, and ${\mathbf D}$ is the {\em diffusion matrix} to be
defined below.  Eq.~(\ref{Eq:diffusion}) should be compared to the
continuous diffusion equation for the particle density: $\partial_t
\rho(\mathbf{r},t) = D\nabla^2\rho(\mathbf{r},t)$, where $D$ is the
diffusion constant. A complex network obviously has an inherent
discreetness, and hence no continuous limit can be taken. However, if
the continuous diffusion equation is understood in its discreet form,
one is lead to regard Eq.~(\ref{Eq:diffusion}) as the {\em master
  equation} for the random-walk process taking place on the underlying
network.  By comparing Eqs.~(\ref{Eq:Cont}) and (\ref{Eq:diffusion}),
as well as taking advantage of $\sum_j A_{ji}\rho_i(t)/k_i =
\rho_i(t)$ (because $\sum_i A_{ij}=k_j$), one is lead to
$D_{ij}=A_{ij}/k_j-\delta_{ij}$. Hence, an equivalent formulation of
Eq.~(\ref{Eq:diffusion}), is
\begin{eqnarray}
  \label{Eq:transfer}
  {\mathbf \rho}(t+1) &=& {\mathbf T} {\mathbf\rho}(t),
\end{eqnarray}
where ${\mathbf T}$ is the {\em transfer matrix}, related to the
diffusion matrix by ${\mathbf T}= {\mathbf D}+{\mathbf 1}$.  In
component form one thus has $T_{ij}=A_{ij}/k_j$. Since the transfer
matrix, ``transfers'' the walker distribution one time step ahead, it
can therefore be thought of as a time-propagator for the process.
Formally the time development, from some arbitrarily chosen initial
state ${\mathbf\rho}(0)$, 
can be obtained by iteration on Eq.~(\ref{Eq:transfer}) with
the result ${\mathbf \rho}(t)={\mathbf T}^t {\mathbf\rho}(0)$.
Here ${\mathbf T}^t$ means the transition matrix to the $t$'th power.

In general, the transfer matrix (or equivalently, the diffusion
matrix) will not be symmetric. However, $\mathbf{T}$ can be related to
a symmetric matrix ${\mathbf S}$ by the following similarity
transformation ${\mathbf S}={\mathbf K}{\mathbf T}{\mathbf K}^{-1}$
with $K_{ij}=\delta_{ij}/\sqrt{k_i}$. Thus, ${\mathbf T}$ and
${\mathbf S}$ will have the same eigenvalue spectrum, and all
eigenvalues $\lambda^{(\alpha)}$ (of $\mathbf{T}$) will be real. It is
this eigenvalue spectrum that will control the time-development of the
diffusive process through $\mathbf{\Lambda}^t$ with
$\Lambda_{ij}=\delta_{ij}\lambda^{(\alpha)}$.  It should be noted that
since the total number of walkers is conserved at all times, one must
have $-1<\lambda^{(\alpha)}\leq 1$, and that at least one eigenvalue
should be one.\footnote{For the diffusion matrix, this means that its
  spectrum, $\{\lambda^{(\alpha)}_D\}$, must satisfy
  $-2<\lambda^{(\alpha)}_D\leq 0$ since
  $\lambda^{(\alpha)}_D=\lambda^{(\alpha)}-1$.}  We will here adopt
the convention and sort the eigenvalues so that $\alpha=1$ corresponds
to the largest eigenvalue $\lambda^{(1)}=1$, $\lambda^{(2)}$ to the
next to largest one, and so on.  Physically the principal eigenvalue
$\lambda^{(1)}=1$, corresponds to a {\em stationary state} where
${\mathbf \rho}(\infty)\propto \mathbf{\rho}^{(1)}$, the diffusive
current flowing from node $i$ to node $j$ is exactly balanced by that
flowing from $j$ to $i$. The stationary state is unique for single
component networks, and is fully determined by the connectivities of
the network according to $\rho_i(\infty)=k_i/(\sum_i k_i)\propto k_i$.
This can be easily checked by substituting this relation into
Eq.~(\ref{Eq:transfer}).  All modes corresponding to eigenvalues
$|\lambda^{(\alpha)}|\neq 1$, are decaying modes since the time
dependence of ${\mathbf \rho}(t)$ enters through $\mathbf{\Lambda}^t$
and one recalls that $|\lambda^{(\alpha)}|\leq 1$.  Notice that
$\lambda^{(\alpha)}>0$ represent non-oscillatory modes, while
$\lambda^{(\alpha)}<0$ correspond to states where oscillations will
take place with time, but this latter possibility will not be
considered here.

The large scale topology of a given complex network reflects itself in
the statistical properties of its diffusion eigenvectors ${\mathbf
  \rho}^{(\alpha)}$.  One such property is the Participation
Ratio~(PR), that will be defined below. It quantifies the effective
number of nodes participating in a given eigenvector with a
significant weight.  Since the stationary state of a network depends
on the connectivities of its nodes, $\rho_i(\infty)\propto k_i$, it is
convenient to introduce a normalized eigenvector
\begin{eqnarray}
c_i^{(\alpha)}(t) &=&  \frac{\rho_i^{(\alpha)}(t)}{k_i}.
\label{currentdef}
\end{eqnarray}
Observe that $c^{(\alpha)}_i$ is nothing but the walker density per
link of node $i$.  Hence, in effect, $c_i^{(\alpha)}$ represents the
{\em outgoing currents} flowing from node $i$, along each of its
links, toward its neighbors.  In the steady state, where
$\rho_i(\infty)\propto k_i$, it follows that these currents are all
the same for any link in the network. Hence, highly connected nodes
are not treated differently from less connected nodes.  More formally,
$\mathbf{c}^{(\alpha)}$ are the eigenvectors of the transposed
transfer matrix $\mathbf{T}^{\dagger}$ corresponding to the same
eigenvalue $\lambda^{(\alpha)}$ (of $\mathbf{T}$).
Here one alternatively could have defined $\mathbf{T}^{\dagger}$
(instead of $\mathbf{T}$) as the transfer matrix from the very
beginning.  Doing so, would have resulted in a master equation of the
form (\ref{Eq:transfer}), but for the currents $\mathbf{c}(t)$ with
corresponding eigenmodes $\mathbf{c}^{(\alpha)}$.  The physical
interpretation of such an alternative equation is as follows: Instead
of walkers, think of a signal propagating on  the
underlying network.  The signal at node $i$ at time $t+1$ is then the
average of the signal at the nearest neighbors at time $t$.

If the link currents are normalized to unity, {\it i.e.} if
$|\!|\mathbf{c}^{(\alpha)}|\!|_2=1$ in the $L_2$-norm, then the
participation ratio~(PR) is defined as~\cite{ParticipationRatio}
\begin{eqnarray}
  \label{eq:PR}
  \chi_{\alpha} &=& \left[\sum_{i=1}^N \left(c^{(\alpha)}_i\right)^{4} \right]^{-1}. 
\end{eqnarray}
For the stationary state, where all the currents are equal to
$c_i^{(1)}=1/\sqrt{N}$, one has $\chi_{1}=N$ where we recall that $N$
is the total number of nodes in the network.  Hence, when $\alpha\neq
1$, the ratio $\chi_{\alpha}$ can be regarded as an effective number
of nodes participating in the eigenvector $\mathbf{c}^{(\alpha)}$ with
a significant weight~\cite{ParticipationRatio,Eriksen_PRL}.  It should
be noted that the participation ratio is a simple and crude measure of
size. Strictly speaking, for this ratio to be able to say something
with confidence about the size of a given module, the main
contribution to $\chi_{\alpha}$ should come from nodes within that
module. For instance, if $\chi_{\alpha}$ gets major contributions from
current elements $c_i^{(\alpha)}$ of different signs, {\it i.e.}  from
different topological structures, it is not trivial to relate
$\chi_{\alpha}$ to the size of a single module.

For all types of complex networks where a modular structure is of
interest, it is important to try to quantify the number of modules.
Such information is of interest since this number may say something
about the organizing principles being active in the generating process
of the network. However, the measurement of this number is often
hampered by statistical uncertainties due to temporary (non-robust)
topological structures being counted as modules.  Hence the challenge
is to obtain the significant (or robust) number of modules that are
due to the ``rule of creation'' and not just happened to be there by
chance. For this purpose, it is useful to introduce a null model --- a
randomized version of the network at hand --- but so that the degree
distribution of the original network is not
changes~\cite{maslov_sneppen_science}.  A rough estimate of the number
of different modules contained in a network could be given by the
number of slowly decaying non-oscillatory modes that have a
participation ratio, $\chi_{\alpha}$, significantly exceeding the
(ensemble) averaged participation ration of the corresponding
randomized network, $\chi_{\alpha}^{\mbox{\small \sc rd}}$. Hence, it
is suggested that a module is significant if $\chi_{\alpha}\gg
\chi_{\alpha}^{\mbox{\small \sc rd}}$.

We will close this section by noting that the (outgoing) currents,
$c_i^{\alpha}$, can also be used for vitalization purposes. The basis
of this approach is the observation that the outgoing currents on
links within a module are almost constant, and the size of this
constant depends on to which extent the module participates in the
given eigenmode. The constants thus varies from module to module and
from eigenmode to eigenmode.  Hence, by sorting $c_i^{(\alpha)}$ by
size, nodes belonging to the same module will be located close to
each.


\section{The data sets}
\label{Sec:data-sets}

During a two year period in the mid 1970's, W.\ Zachary studies the
relations among $34$ members of a university karate club during a
period of trouble~\cite{Zachary}. A serious controversy existed
between the trainer of the club~(node $1$ in Fig.~\ref{Fig:Zachary}a)
and its administrator~(node $34$).  Ultimately this conflict resulted
in the breakup of the club into two new clubs of roughly the same
size. In his original study~\cite{Zachary}, W.\ Zachary mapped out the
strength of friendship between the various members. In our study,
however, we will be considering the unweighted version of his
network~\cite{Download} (see Fig.~\ref{Fig:Zachary}a). Recently, this
same network was used by Girvan and Newman~\cite{Girvan2002} in their
study of community structures.

The second network that will be considered is a much larger network
taken from the organization of the Internet.  The Internet consists of
a large number of individual computers that are identified by their
so-called IP-address, and they are usually grouped together in local
area networks~(LAN). Such computer networks are connected to one
another via routers --- a complex network linking device. In addition
to being able to direct pieces of information to its intended
destination, a router also has the ability to determine the best path
to a given destination (routing). This is done by keeping available an
updated routing table telling the router how to reach certain
destinations specified by the network administrator.  In much the same
way as computers are being organized into networks, the same is done
for routers.  If the router network becomes large and coherent enough,
it may make out what is called an {\em Autonomous System}~(AS). An AS
is a connected segment of a network that consists of a collection of
subnetworks (with hosts attached) interconnected by a set of routes.
Usually it is required that the subnetworks and the routers should be
controlled by a coherent organization --- like, say, a university, or
a medium-to-big enterprise.  Importantly for the efficiency of the
Internet, each Autonomous System, identified by a unique AS number, is
expected to present to the other systems a consistent list of
destinations reachable through that AS, as well as a coherent interior
routing plan.

The particular network we will be considering is an Autonomous System
network collected by the Oregon Views project on January 3,
2000~\cite{AS}.
In this network the AS will act as nodes, while their routing plans,
{\it i.e.} with which other AS a given one directly shares
information, will correspond to the links of the network.


\section{Results and discussion}
\label{Sec:Results}

\subsection{The Zachary Friendship Network}

\begin{figure}[htbp]
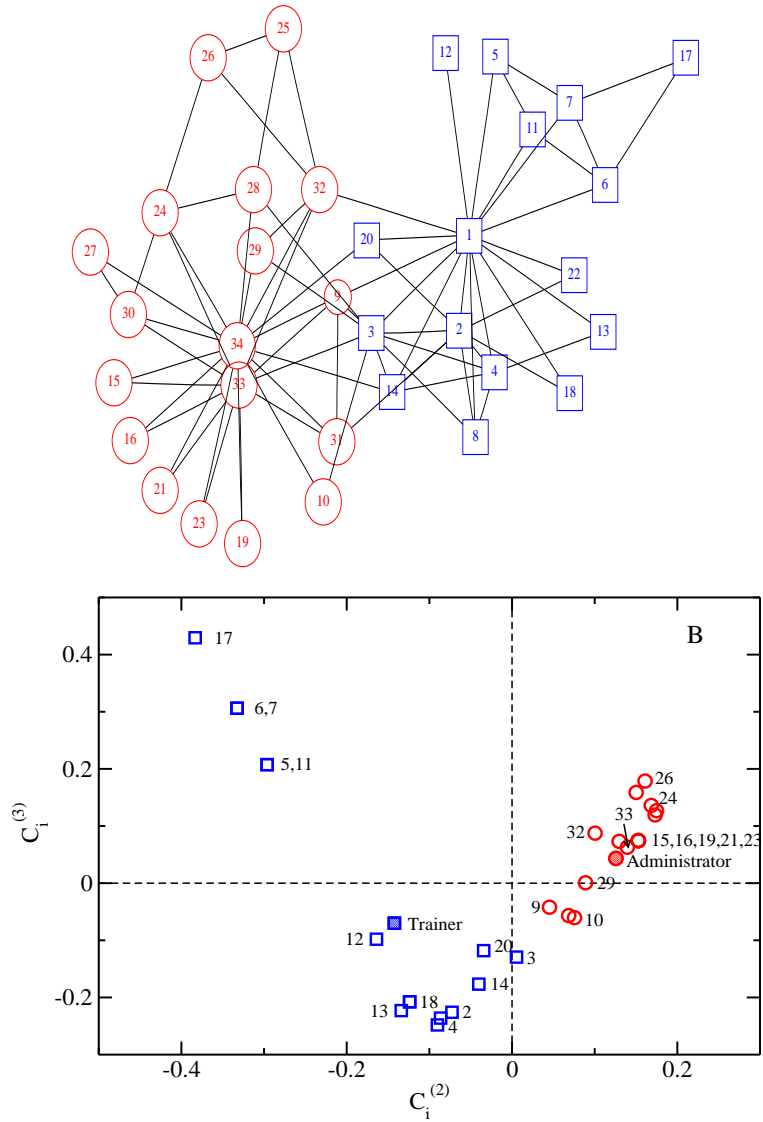

  \centering
    \includegraphics*[width=10cm,height=8cm]{zachary_network}\\
    \includegraphics*[width=10cm,height=7cm]{current} 
  \caption{(a) Zachary's friendship network~\protect\cite{Zachary} 
    of the ``troubled'' karate club consisting of $N=34$ nodes and
    $L=78$ links (Figure after Ref.~\protect\cite{Girvan2002}).  Here
    open squares and circles are used to denote the supporters, in the
    ongoing conflict, of the trainer (node $1$) and administrator
    (node $34$), respectively. (b) The $c^{(2)}c^{(3)}$-plot maps out
    the large scale topology of the network.
    The dashed lines indicate the lines of zero currents. }
  \label{Fig:Zachary}
\end{figure}

In Sec.~\ref{Sec:Method}, it was argued that from the currents
corresponding to the {\em slowly} decaying eigenmodes, one should be
able map out the large scale structure of the underlying network. Our
hypothesis will now be put to the test, and we start with the small
friendship network due to Zachary~\cite{Zachary}.  This network is
depicted in Fig.~\ref{Fig:Zachary}a, and it is small enough to enable
us to see how it all comes about.  By visual inspection of this
figure, it is apparent that there are, at least, two large scale
clusters --- one corresponding to the trainer of the karate club~(node
$1$) and his supporters, and one to the administrator~(node $34$).  In
Fig.~\ref{Fig:Zachary}a, the karate club members (according to
Zachary) supporting the trainer in the ongoing conflict are marked
with squares ($16$ nodes), while the followers of the administrator
are marked with open circles ($18$ nodes).  Notice that this
subdivision was done by Zachary~\cite{Zachary}, and no sophisticated
clustering techniques were applied to obtain these results.

We will now see if the same clustering structure can be obtained by
the method outlined in Sec.~\ref{Sec:Method} of this paper.  Let us
start by considering the two most slowly decaying modes of the
network, namely $\alpha=2$ and $3$ (with our ordering), corresponding
to the eigenvalues $\lambda^{(2)}= 0.87$ and $\lambda^{(3)}= 0.71$,
respectively.  If one sorts the elements of the current vector
$\mathbf{c}^{(2)}$ by size, and group them according to their signs,
one recovers two clusters (see the abscissa of
Fig.~\ref{Fig:Zachary}b).  This grouping, or topological structure,
fits nicely with the original assignments made by
Zachary~(Fig.~\ref{Fig:Zachary}a) and indicated by the open squares
and circles in Fig.~\ref{Fig:Zachary}b. Hence it is the
trainer-administrator separation of the karate club members that is
mapped out by the $\mathbf{c}^{(2)}$-currents. Observe that node~$3$,
which has an equal number of links to supporters of the trainer and
administrator, has a current that is almost zero. Our identification
of this node is the only one that differs from that made originally by
Zachary. Interestingly, our classification, including node~$3$, fits
that previously obtained by Girvan and Newman~\cite{Girvan2002} in
their hierarchical tree clustering approach.

If the same procedure was repeated, but for the currents
$\mathbf{c}^{(3)}$, one would map out other topological features of
the network. In Fig.~\ref{Fig:Zachary}b, a $c^{(2)}c^{(3)}$-plot is
presented. Such a plot groups the nodes along two axis; the
$c^{(2)}$-axis --- corresponding to the trainer-administrator axis ---
and the $c^{(3)}$-axis. The most striking feature of
Fig.~\ref{Fig:Zachary}b is the group of nodes corresponding to
$c^{(2)}<0$ and $c^{(3)}>0$, {\it i.e.} to the following group of
nodes $\{5,6,7,11,17\}$.  In Fig.~\ref{Fig:Zachary}a, these nodes are
located in the upper left corner of the graph. They are in addition to
being connected among themselves, only connected to the rest of the
network via the trainer.  Thus, they represent a sub-cluster, and this
is indeed what is apparent from Fig.~\ref{Fig:Zachary}b.

The participation ratios for the two slowest decaying diffusive modes
were found to be $\chi_2=14.0$ and $\chi_3=13.1$, respectively. As can
be seen from Fig.~\ref{Fig:Zachary}b, these participation ratios
receive substantial contributions from both positive and negative
current elements, {\it i.e.} from more then one topological structure.
Hence, we suspect that these numbers being close to the size of the
administrator and trainer ``clan'' (of size $18$ and $16$ nodes,
respectively) is somewhat accidental.

\subsection{The Autonomous System Network}

We will now focus our attention on the Autonomous System~(AS) network
introduced in the previous Section.  This network, consisting of
$N=6,474$ nodes and $12,572$ undirected links, is so big that plotting
it for the purpose of study its modular structure is not a practical
option.  Motivated by what we found for the small Zachary friendship
network, we will now go ahead and use somewhat similar techniques for
its study.

\begin{figure}[bp]
  \centering
  \includegraphics*[width=10cm,height=7cm]{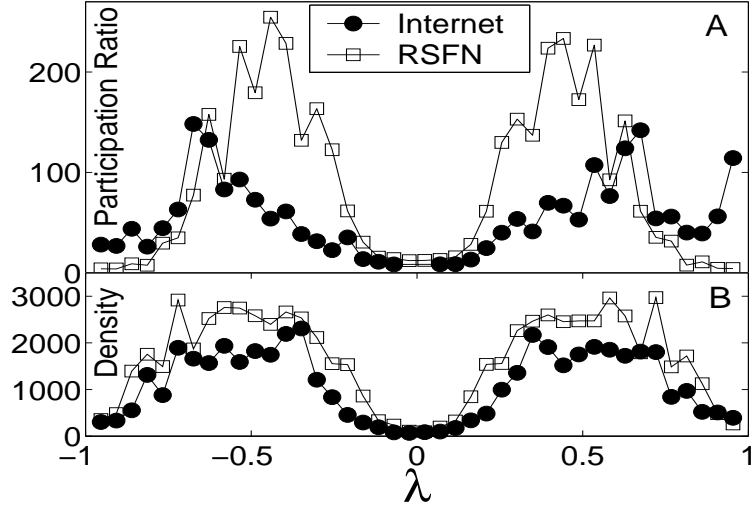}
  \caption{
    The participation ratio $\chi_{\alpha}$ (top, A) and the
    eigenvalue density (bottom, B) as a function of the eigenvalue of
    the transfer matrix, $-1<\lambda^{(\alpha)}<1$, measured in the
    Internet (filled circles) and in its randomized counterpart (open
    squares) --- a Random Scale-Free Network. The participation ratio
    was averaged over $\lambda$-bins of size 0.05 excluding eigenmodes
    $\lambda^{(\alpha)}=0$, and $\lambda^{(1)}=1$ (cf.
    Ref.~\cite{{Eriksen_PRL}}).
  }
\label{Fig:Internet}
\end{figure}

\begin{figure}[tbh]
    \centering
    \includegraphics*[width=10cm,height=8cm]{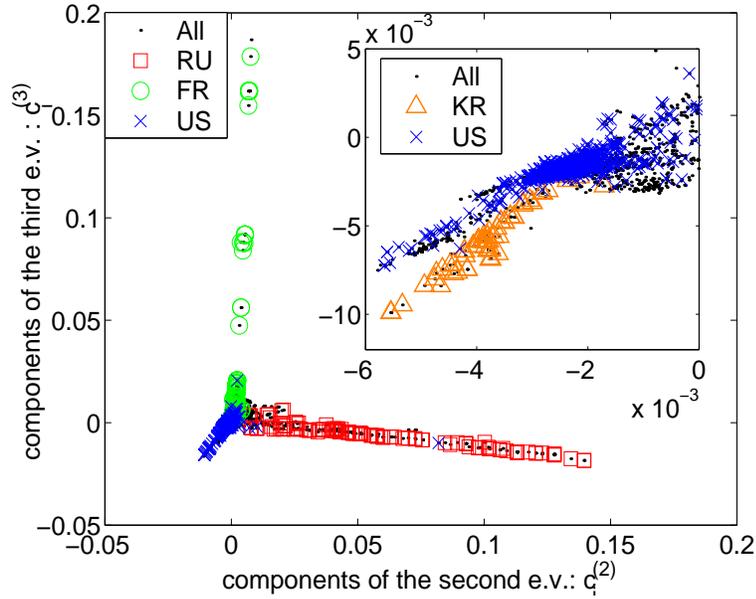}
    \caption{The Internet clustering:
      The coordinates of the $i$-th AS in this plot are its current
      components $(c_i^{(2)},c_i^{(3)})$ of the two slowest decaying
      non-oscillatory diffusion modes. The symbols reveals the
      geographical location of the AS: Russia -- squares, France --
      circles, USA -- crosses, Korea -- triangles). Note the straight
      lines corresponding to good country-modules.  }
  \label{Fig:Internet-Currents}
  \end{figure}

In Fig.~\ref{Fig:Internet} the participation ratio, $\chi_{\alpha}$,
of eigenvectors $c_i^{(\alpha)}$ (top) and the eigenvalue density
(bottom) are plotted as functions of the corresponding eigenvalues
$-1<\lambda_i^{(\alpha)}<1$.  The data for the Internet, that is an
example of a Scale-Free Network~ \cite{Faloustos}, are displayed
together with the data for its randomized counterpart~(the null
model).  From Fig.~\ref{Fig:Internet} it is apparent that while the
density of states is rather similar for these two networks, the
participation ratios of the slowly decaying modes, especially for
$\lambda^{(\alpha)}$ close to $1$, are markedly higher in the Internet
network than in the accompanying randomized network. These differences
signal large scale topological structures~\cite{Eriksen_PRL} that are
real and not accidental. For the most slowly decaying diffusive
eigenmode, the participation ratio is $\chi_2\simeq 100$ (cf.
Fig.~\ref{Fig:Internet}a).  From Fig.~\ref{Fig:Internet-Currents},
that depicts the currents of the two slowest decaying diffusive
eigenmodes, one observes that the main contribution to $\chi_{2}$
comes from current elements $c_i^{(2)}$ of the same sign. Thus, a
module has been detected, and $\chi_{2}$ roughly measures its size.
From, Fig.~\ref{Fig:Internet-Currents} one also makes the interesting
observation that the main contributing nodes to $\chi_{2}$ are
Autonomous Systems located in Russia (denoted by open squares in
Fig.~\ref{Fig:Internet-Currents}). Hence, the diffusive mode
$\alpha=2$ maps out Russia, as a Russian module~\cite{Eriksen_PRL}!
In total there are $174$ Russian nodes in our data set.  Moreover,
from Fig.~\ref{Fig:Internet-Currents}, modules corresponding to
countries like the USA, France and Korea are also easily identified.
The edges of the AS Internet network, defined as the nodes
corresponding to the most distant values of the current elements
$c_i^{(2)}$, are in this case represented by a Russian and a US
military site located in the South Pacific. These nodes can thus be
said to represent the extreme edges of the Internet.
One has also studies the number of significant modules (where
$\chi_{\alpha} \gg \chi_{\alpha}^{\mbox{\small \sc rd}}$), and found
that this number for the AS network is roughly
$M=100$~\cite{Eriksen_PRL}.  The number of different nodes
participating in one or more of these $M$ modules was found to be
about $N_M\sim 1800$. Hence the modularity of the network is (at
least) $N_M/N\simeq 30\%$.

\subsection{Generic features of the current-current plots}

One common feature of the current-current plots,
Figs.~\ref{Fig:Zachary}b and \ref{Fig:Internet-Currents}, is their
line or star-like structures. Such structures are in fact rather
generic, and we will here try to explain why. We have argued earlier
that current elements $c_i^{(\alpha)}$ should almost be the same
within a module.  Hence, for two nodes $i$ and $j$ belonging to one
and the same module, the fraction $c_i^{(\alpha)}/c_j^{(\alpha)}$ is
expected to be a constant unique for that module, but independent of
the diffusive mode $\alpha$. For two different (significant) diffusive
modes, say $\alpha$ and $\beta$, the fraction
$\gamma=c_i^{(\alpha)}/c_i^{(\beta)}$ (for node $i$) will therefore as
a consequence also be constant. Thus, under the assumptions made
above, one predicts a straight line in a current-current plot (that
pases through the origin)~\cite{Eriksen_PRL} :
$c_i^{(\alpha)}=\gamma\,c_i^{(\beta)}$, for nodes $i$ belonging to the
same module.  From Fig.~\ref{Fig:Internet-Currents}, this simple
argument appears to be valid. That Fig.~\ref{Fig:Zachary}b seems to
not pass through the origin, but still being straight lines, we
believe is due to the diffusive modes being excited over a non-trivial
background in this case.

\subsection{Numerical implementation}

Before presenting the conclusions of this paper, we will add a few
closing remarks on the numerical implementation of the method.  For
this work to have any relevance for large real-world networks, a fast,
memory saving, and optimized algorithm for the calculation of the
largest eigenvalues and the corresponding eigenvectors of a sparse
matrix is required.  Fortunately, such an algorithm has already been
implemented and made available {\it e.g.} through the TRLan software
package~\cite{TRLan}.  This software is optimized for handling large
problems, and it can run on large scale parallel supercomputers.


\section{Conclusions}
\label{Sec:Conclusions}

We have generalized the normal diffusion process to diffusion on
(discreet) complex networks. By considering such a process, it has
been demonstrated that topological properties --- like modular
structures and edges --- of the underlying network can be probed. This
is achieved by focusing on the slowest decaying eigenmodes of the
network. The use of the procedure was exemplified by considering a
small friendship network, with known modular structure, as well as a
routing network of the Internet, where the structure was not so well
known.  For the friendship network the known structure was well
reproduced, and the Internet was indeed found to be modular.  The
detected modules of the Internet were consistent with the geographical
location of the nodes, and the individual modules corresponded roughly
to the national structure.  Interestingly, it was observed that a
political subdivision of the Internet was also one of the predictions
of the algorithm presented in this paper; The two most poorly connected
nodes of the Internet (extreme edges), where found to be represented
by a Russian and a US military site located in the South Pacific.


\section*{Acknowledgment}

Work at Brookhaven National Laboratory was carried out under Contract
No. DE-AC02-98CH10886, Division of Material Science, U.S.\ Department of
Energy.



\end{document}